\documentclass{article}

\usepackage{PRIMEarxiv}

\usepackage[utf8]{inputenc} 
\usepackage[T1]{fontenc}    
\usepackage{hyperref}       
\usepackage{url}            
\usepackage{booktabs}       
\usepackage{amsmath}
\usepackage{amsthm}
\usepackage{amssymb}
\usepackage{amsfonts}
\usepackage{mathtools}
\usepackage{amsfonts}       
\usepackage{nicefrac}       
\usepackage{microtype}      
\usepackage{fancyhdr}       
\usepackage{graphicx}       
\graphicspath{{media/}}     
\usepackage{times}
\usepackage{subfig}
\usepackage{xcolor}
\usepackage{array}
\usepackage{cite}
\usepackage{multirow}
\usepackage{adjustbox}
\usepackage[scr=dutchcal]{mathalfa}
\usepackage[font=small,labelfont=bf]{caption}
\DeclareMathOperator{\sign}{sign}
\DeclareMathOperator{\diag}{diag}
\DeclareMathOperator{\tr}{tr}

\newtheorem{theorem}{Theorem}

\newtheorem{assumption}{Assumption}

\newtheorem{remark}{Remark}

\usepackage{tikz-network}
\tikzset{
    block/.style = {draw, fill=white, rectangle, minimum height=.75cm, minimum width=1.25cm},
    tmp/.style  = {coordinate}, 
    sum/.style= {draw, fill=white, circle},
    input/.style = {coordinate},
    output/.style= {coordinate},
    pinstyle/.style = {pin edge={to-,thin,black}
    }
}

\title{Model Reference Adaptive Control of Networked Systems with State and Input Delays
\thanks{This research is fully supported by Department of Engineering Physics, Institut Teknologi Sepuluh Nopember, under grant No.1868/PKS/ITS/2023.} 
}

\author{
  Moh. Kamalul Wafi, Katherin Indriawati, Bambang L. Widjiantoro \\
  Department of Engineering Physics \\
  Institut Teknologi Sepuluh Nopember \\
  Surabaya, Indonesia \\
  \texttt{\{kamalul.wafi\}@its.ac.id} \\
}

\begin{document}
\maketitle

\begin{abstract}
Adaptive control strategies have progressively advanced to accommodate increasingly uncertain, delayed, and interconnected systems. This paper addresses the model reference adaptive control (MRAC) of networked, heterogeneous, and unknown dynamical agents subject to both state and input delays. The objective is to ensure that all follower agents asymptotically track the trajectory of a stable leader system, despite system uncertainties and communication constraints. Two communication topologies are considered: (i) full connectivity between each agent and the leader, and (ii) partial connectivity wherein agents rely on both neighboring peers and the leader. The agent-to-agent and agent-to-leader interactions are encoded using a Laplacian-like matrix and a diagonal model-weighting matrix, respectively. To compensate for the delays, a predictor-based control structure and an auxiliary dynamic system are proposed. The control framework includes distributed adaptive parameter laws derived via Lyapunov-based analysis, ensuring convergence of the augmented tracking error. Stability conditions are established through a carefully constructed Lyapunov–Krasovskii functional, under minimal assumptions on connectivity and excitation. Numerical simulations of both network structures validate the proposed method, demonstrating that exact leader tracking is achieved under appropriately designed learning rates and initializations. This work lays a foundation for future studies on fault-resilient distributed adaptive control incorporating data-driven or reinforcement learning techniques.
\end{abstract}

\allowdisplaybreaks
 
\section{Introduction}\label{S1}
With the advancement of increasingly complex systems, the development of control methodologies has seen significant evolution, ranging from classical and modern techniques to learning-based algorithms. Among these, adaptive control has emerged as a powerful approach for handling uncertain systems while maintaining desired performance conditions \cite{R1}. However, adaptive control still faces persistent challenges. These include stability analysis in model-free contexts \cite{R8,R7}, handling time-varying parameterizations \cite{R10}, achieving asymptotic tracking \cite{R11}, and rejecting disturbances with time-varying frequencies \cite{R16}. Other issues involve limited frequency-domain Linear Matrix Inequality (LMI) certifications \cite{R18}, and the scaling and estimation complexities that arise in non-minimum phase systems, particularly in microgrid applications \cite{R19,R20,R21,R23}.

Some researchers have attempted to simplify these problems by converting them into classical formulations for approximation \cite{R24}, or by utilizing iteration-based adaptive control schemes to circumvent the need for state estimation \cite{R25}. Despite these developments, a significant gap remains in handling model reference adaptive control (MRAC) systems with unknown, distributed (i.e., non-lumped) delays \cite{R26,R27,R28,R29}. These distributed delays present a considerable challenge, especially in networked environments where delays are not centralized or deterministic. To address this issue, several strategies have been proposed, including dynamic switching schemes \cite{R12,R13}, and MRAC formulations for multivariable systems with uncertain time delays \cite{R14,R15,R3}, accounting for time-varying control bounds and exogenous disturbances. Among the classical MRAC design techniques, notable methods include the Lyapunov-based approach \cite{R32,R33}, the MIT rule \cite{R34}, and the adaptation-based strategy \cite{R35}. In this work, we adopt the approach introduced in \cite{R35} to handle delays effectively, particularly in practical applications where communication signals are limited.

Specifically, this paper focuses on direct MRAC for linear systems affected by both input and state delays. Our proposed construction relies on predicting the reference trajectory and formulating an augmented system, using Lyapunov–Krasovskii-based stability updates. The problem is further extended to a networked multi-agent setting \cite{R46,R47,R48}, where a collection of unknown and potentially unstable agents aim to follow a stable reference (or leader) despite communication constraints and known, lumped delays. Two scenarios are explored in this context: one where each agent has full access to the leader’s signal, and another where agents communicate with both the leader and their immediate neighbors. The communication topology is characterized using a Laplacian-like matrix, while the leader-agent interactions are weighted by a model-dependent weight matrix. To reflect real-world limitations, both state and input delays are incorporated into the network dynamics. These delays are addressed via reference signal prediction and system augmentation techniques designed to ensure convergence and stability. Looking ahead, future research will build on recent developments in adaptive learning-based control \cite{R2,R4,R5}, focusing on robust and intelligent systems that can adapt to nonlinearities and uncertain environments. After a detailed analysis of system performance and delay sensitivity, our long-term goal is to develop distributed adaptive and fault-tolerant control strategies for nonlinear networked systems operating under complex conditions \cite{R9,R36,R37,R38,R41,R42,R43}.

The remainder of this paper is organized as follows: Section \ref{S2} formulates the problem of adaptive control for networked systems with state and input delays. Section \ref{S3} presents the adaptation approach and stability analysis. Section \ref{S4} provides numerical examples for both network scenarios to evaluate the proposed method's performance. Finally, Section \ref{S5} concludes the paper and outlines potential directions for future research. 

\textbf{Notations.}
Let $\mathbb{R}^p$ denote the $p$-dimensional real Euclidean space, and $\mathbb{R}^{p \times q}$ the set of real matrices of dimension $p \times q$. The identity matrix of size $p$ is denoted by $I_p \in \mathbb{R}^{p \times p}$, while $\mathbf{1}_p=[1,\dots,1]^\top \in \mathbb{R}^p$ and $\mathbf{0}_p=[0,\dots,0]^\top \in \mathbb{R}^p$ represent the vectors of all ones and all zeros, respectively. A diagonal matrix with entries $p_i$ is written as $P = \diag\{p_i\}$. The Kronecker product between two matrices $A$ and $B$ is denoted by $A \otimes B$, while $A \odot B$ indicates their Hadamard (element-wise) product. The trace of a square matrix $P$ is denoted by $\tr[P]$, and $|P|$ represents the matrix of element-wise absolute values of $P$. Matrix norms used throughout the paper include the Euclidean (or spectral) norm $\lVert P \rVert_2$, the Frobenius norm $\lVert P \rVert_F$, and the element-wise $\ell_1$ norm $\lVert P \rVert_1 = \sum_{i,j} |P_{ij}|$ when necessary. For vectors $x$, we similarly define the $\ell_2$ norm as $\lVert x \rVert_2 = \sqrt{x^\top x}$. Time-varying functions are denoted explicitly with $(t)$, e.g., $x(t)$, and delayed signals are written as $x(t - \tau)$ where $\tau \in \mathbb{R}_{\geq 0}$ is a constant delay. The function $f(t+\tau|t)$ denotes a prediction of $f$ at time $t+\tau$ based on data available up to time $t$. For compactness in block matrix expressions, we denote block-diagonal matrices as $\diag\{M_1, \dots, M_\ell\}$, and use stacked vectors such as $\bar{x} = [x_1^\top, \dots, x_\ell^\top]^\top \in \mathbb{R}^{\bar{n}}$ where $\bar{n} = n \times \ell$.

\section{Adaptive Control of Networked Systems}\label{S2}
\begin{figure}[t!]
    \centering
    \subfloat[\label{F1a} Example 1]{
        \includegraphics[width=0.25\textwidth]{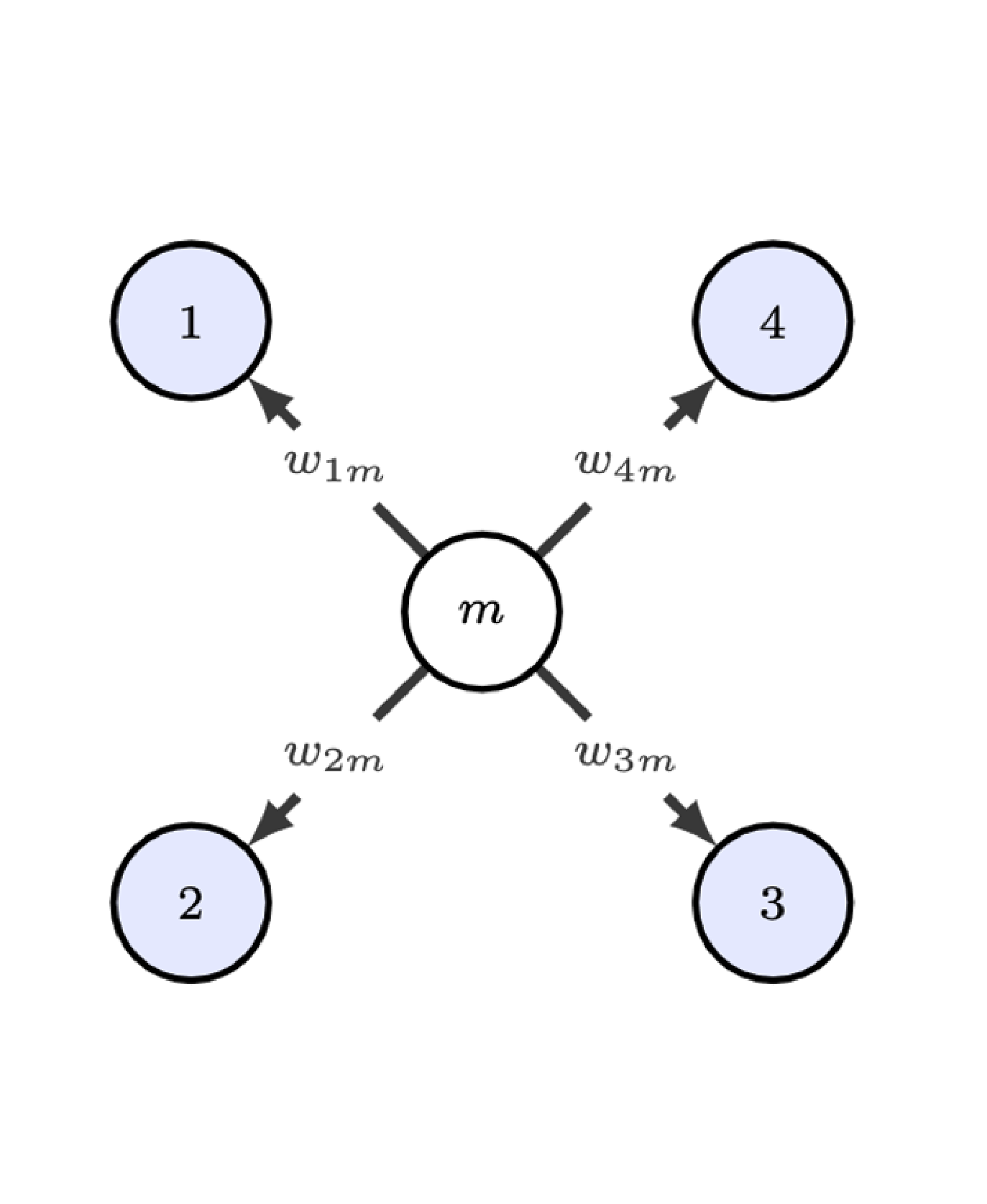}}
    \subfloat[\label{F1b} Example 2]{
        \includegraphics[width=0.25\textwidth]{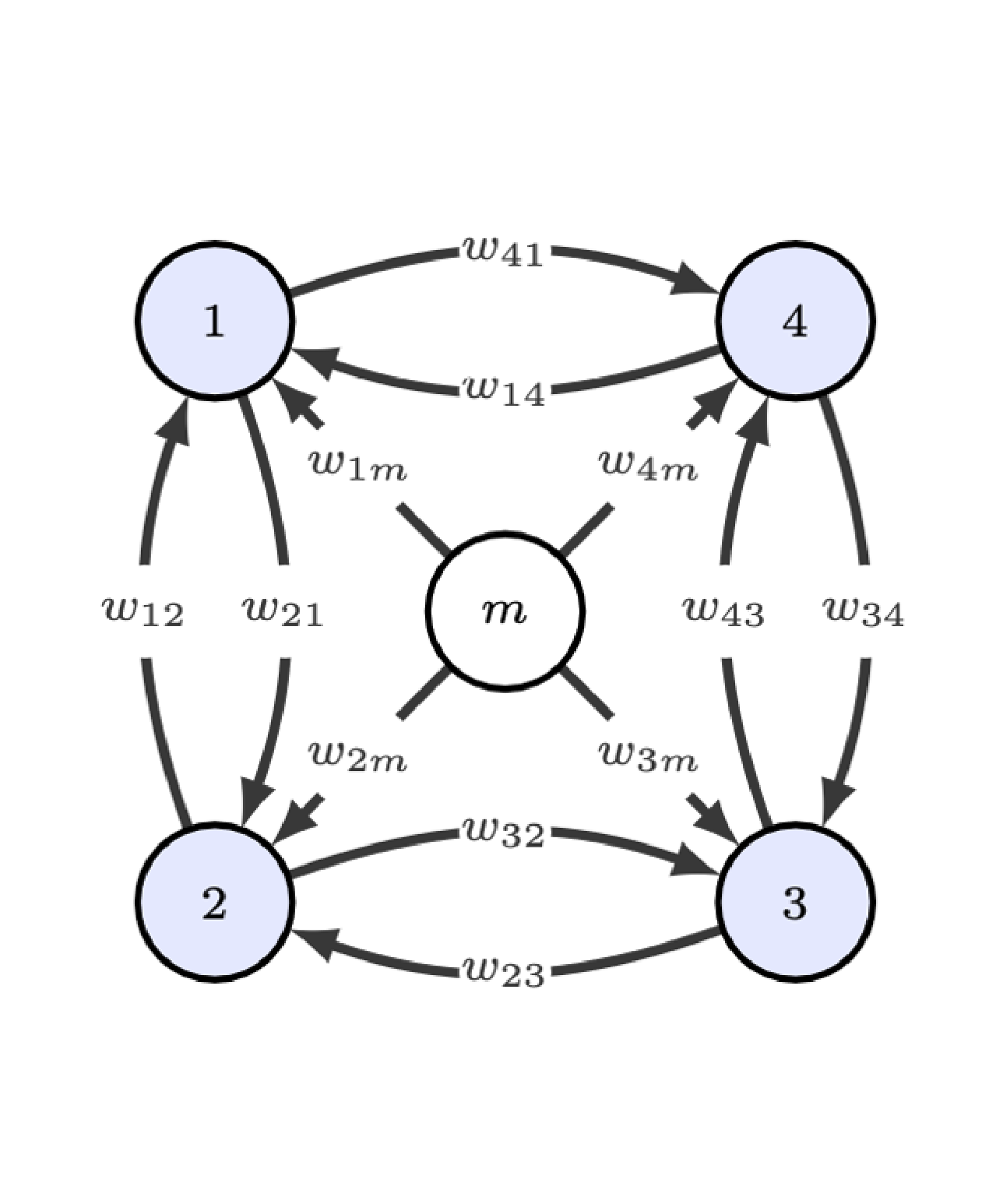}}
    \subfloat[\label{F1c} Block diagram]{
        \includegraphics[width=0.495\textwidth]{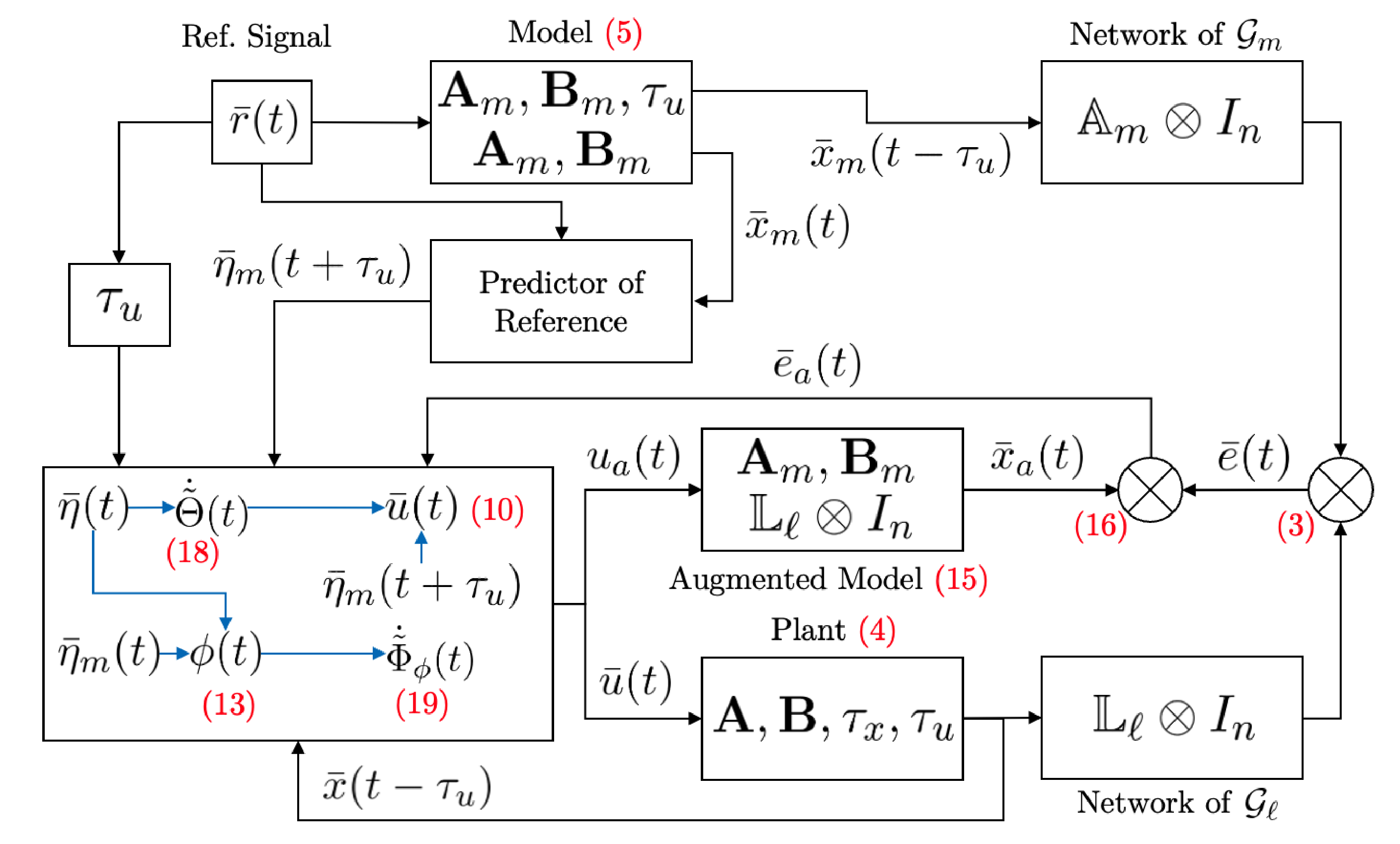}}
    \caption{\small The two network scenarios shown in (a) and (b) while (c) defines the block diagram of the proposed method.}
    \label{F1}
\end{figure}

We consider a networked system consisting of $\ell$ uncertain linear agents and a leader (or model reference), denoted by the subscript $m$, where both state and input delays are present. Let $\tau_x$ and $\tau_u$ represent the state and input delays, respectively, with $\tau_x \leq \tau_u$. The system structure is illustrated in Fig.~\ref{F1a} and Fig.~\ref{F1b}. Each of the $\ell$ agents is governed by the following delayed differential equation:
\begin{align}
    \dot{x}_i(t) = A_i x_i(t) - A^\zeta_i x_i(t - \tau_x) + B_i u_i(t - \tau_u) \label{E1}
\end{align}
where $x_i \in \mathbb{R}^{n}$ and $u_i \in \mathbb{R}^{p}$ denote the state and control input of the $i$-th agent, for $i = 1, \dots, \ell$. The matrices $A_i \in \mathbb{R}^{n \times n}$, $B_i \in \mathbb{R}^{n \times p}$, and $A^\zeta_i \in A_i$ are constant but uncertain real matrices. The leader system, serving as the desired model reference, is defined as:
\begin{align}
    \dot{x}_m(t) = A_m x_m(t) + B_m r(t - \tau_u) \label{E2}
\end{align}
where $x_m \in \mathbb{R}^{n}$ is the state of the leader and $r \in \mathbb{R}^{p}$ is the reference input signal. The matrices $A_m \in \mathbb{R}^{n \times n}$ and $B_m \in \mathbb{R}^{n \times p}$ are known and constant. The objective of this study is to design distributed local control laws $u_i(t)$ for each agent such that the state trajectories of the agent subsystems \eqref{E1} asymptotically follow the reference trajectory defined by the leader system in \eqref{E2}, despite the presence of communication-induced state and input delays and system uncertainties.

We model the communication topology among the $\ell$ agents and the leader using a weighted directed graph (digraph) defined as $\mathcal{G} \coloneqq \left\{ \mathcal{V} = \{1, \dots, \ell\} \cup \{m\},\ \mathcal{E},\ w_{ij} \right\}$
where:
\begin{itemize}
    \item $\mathcal{V}$ is the set of nodes, consisting of the $\ell$ agents and the leader $m$,
    \item $\mathcal{E} \subseteq \mathcal{V} \times \mathcal{V}$ is the set of directed edges, and  
    \item $w_{ij} > 0$ is the weight associated with the edge from agent $j$ to agent $i$; if no such edge exists, then $w_{ij} = 0$.
\end{itemize}

The directed nature of $\mathcal{G}$ implies that communication is not necessarily bidirectional; that is, $(i,j) \in \mathcal{E}$ does not imply $(j,i) \in \mathcal{E}$. For each agent $i \in \{1, \dots, \ell\}$, we define its neighborhood $\mathcal{N}_i$ as the set of agents $j$ from which it receives information, i.e., $\mathcal{N}_i = \{j \in \mathcal{V} \mid (i,j) \in \mathcal{E}\}$.

To facilitate the analysis, we partition the graph $\mathcal{G}$ into two induced subgraphs:
\begin{itemize}
    \item $\mathcal{G}_\ell = \{\mathcal{V}_\ell, \mathcal{E}_\ell, \mathcal{W}_\ell\}$ — capturing the interaction among the $\ell$ follower agents,
    \item $\mathcal{G}_m = \{\mathcal{V}_m, \mathcal{E}_m, \mathcal{W}_m\}$ — modeling the directed connections from the leader $m$ to the followers.
\end{itemize}
This decomposition allows for the separation of leader-follower dynamics from the internal interactions among followers. 

In this framework, we assume that the network is \textit{balanced}, meaning that for each agent $i$, the sum of the incoming edge weights satisfies $w_i = \sum_{j\in\mathcal{N}_i} w_{ij} = 1.$
As a result, the degree matrix of the agent network is the identity matrix, i.e., $\mathbb{D} \coloneqq \diag\{w_1, \dots, w_\ell\} = I_\ell$ and the \textit{Laplacian-like matrix} $\mathbb{L}_\ell$ is defined as $\mathbb{L}_\ell \coloneqq \mathbb{D} - \mathbb{A}_\ell,$ where $\mathbb{A}_\ell$ is the adjacency matrix corresponding to $\mathcal{G}_\ell$ such that $\mathbb{A}_\ell(i,j) = w_{ij}$ if $(i,j) \in \mathcal{E}_\ell$, and $0$ otherwise. Additionally, we define $\mathbb{A}_m \in \mathbb{R}^{\ell \times \ell}$ as a diagonal matrix capturing the direct influence (weights) of the leader on each agent, with non-zero entries only for those agents directly connected to the leader. That is $\mathbb{A}_m \coloneqq \diag\{w_{1m}, \dots, w_{\ell m}\}$ such that $\mathbb{A}_m(i,i) = w_{im}$ if $(i,m) \in \mathcal{E}_m$, and $0$ otherwise.

\begin{remark}[Threshold of Network]\label{rem1}
    To ensure sufficient coupling among the agents and between agents and the leader, we assume that the smallest non-zero eigenvalues of both $\mathbb{L}_\ell$ and $\mathbb{A}_m$ are bounded below by a constant $\vartheta > 0$. That is, $\lambda_i^\ell \geq \vartheta$ and $\lambda_i^m \geq \vartheta$, with a typical design choice being $\vartheta = 0.1$. This threshold reflects the minimum required connectivity strength and is user-defined, depending on the network’s complexity.
\end{remark}
\begin{remark}[Communication Network Properties]\label{rem2}
    The overall network is designed to be balanced in the sense that  
    \begin{equation*}
        \left[ (\mathbb{L}_\ell - \mathbb{A}_m) \otimes I_n \right] \mathbf{1}_{\bar{n}} = 0,
    \end{equation*}
    This condition guarantees consistent information flow across the network. Furthermore, the network topology must include a \textit{directed path from the leader node} $\mathcal{G}_m$ \textit{to every agent node} in $\mathcal{G}_\ell$, as illustrated in Fig.~\ref{F1a} and Fig.~\ref{F1b}. If such paths do not exist, then some eigenvalues $\lambda_i^m$ may drop below the threshold $\vartheta$, i.e. $\exists\lambda_i^m=0<\vartheta$ for some $i$, violating Assumption~\ref{rem1}. In such cases, the design of the augmented leader system, e.g., as defined in Eq.~\eqref{E15}, must be modified to restore proper connectivity.
\end{remark}

We define the synchronization error for each agent $i$ as $e_i \in \mathbb{R}^n$, which quantifies the deviation of agent $i$ from its neighbors and the leader. Using the collective agent state vector $\bar{x}$ and the leader state vector $\bar{x}_m = \mathbf{1}_\ell \otimes x_m$, the error is expressed as:
\begin{align}
    \begin{aligned}
        e_i &\coloneqq \left[ (\mathbb{L}_\ell \otimes I_n)\bar{x} - (\mathbb{A}_m \otimes I_n)\bar{x}_m \right]_i \\
        &= (w_i \otimes I_n)x_i - \sum_{j \in \mathcal{N}_i} (w_{ij} \otimes I_n) x_j,
    \end{aligned}
    \quad\longrightarrow\quad
    \mathbb{L}_\ell =
    \begin{bmatrix}
        w_1 & \cdots & -w_{1\ell} \\
        \vdots & \ddots & \vdots \\
        -w_{\ell 1} & \cdots & w_{\ell}
    \end{bmatrix}, \quad
    \mathbb{A}_m =
    \begin{bmatrix}
        w_{1m} & \cdots & 0 \\
        \vdots & \ddots & \vdots \\
        0 & \cdots & w_{\ell m}
    \end{bmatrix}
    \label{E3}
\end{align}
The overall synchronization error vector is denoted by $\bar{e} = [e_1^\top, \dots, e_\ell^\top]^\top \in \mathbb{R}^{\bar{n}}$, and the control objective is to design a distributed controller for each agent such that $\lim_{t \to \infty} \bar{e}(t) \to 0,$ ensuring that all follower agents asymptotically track the leader's trajectory.

We conclude the problem formulation by describing the compact form of the interconnected dynamics in \eqref{E1}, expressed as:
\begin{align}
    \dot{\bar{x}}(t) = \mathbf{A}\bar{x}(t) + \mathbf{A}^\zeta \bar{x}(t - \tau_x) + \mathbf{B} \bar{u}(t - \tau_u) \label{E4}
\end{align}
where $\bar{x} = [x_1^\top, \dots, x_\ell^\top]^\top \in \mathbb{R}^{\bar{n}}$ with $\bar{n} = n \times \ell$ is the global state vector, and $\bar{u} = [u_1^\top, \dots, u_\ell^\top]^\top \in \mathbb{R}^{\ell}$ is the stacked input vector. The block-diagonal matrices of the multi-agent systems are defined as $\mathbf{A} \coloneqq \diag\{A_1, \dots, A_\ell\} \in \mathbb{R}^{\bar{n} \times \bar{n}}$, $\mathbf{B} \coloneqq \diag\{B_1, \dots, B_\ell\} \in \mathbb{R}^{\bar{n} \times \bar{p}}$ with $\bar{p} = p\times\ell$, and $\mathbf{A}^\zeta \coloneqq \diag\{ A^\zeta_1, \dots, A^\zeta_\ell \}\in \mathbb{R}^{\bar{n} \times \bar{n}}$.

The leader dynamics in \eqref{E2} is also extended into a compact form using block matrices:
\begin{align}
    \dot{\bar{x}}_m(t) = \mathbf{A}_m \bar{x}_m(t) + \mathbf{B}_m \bar{r}(t - \tau_u) \label{E5}
\end{align}
where $\bar{x}_m = \mathbf{1}_\ell \otimes x_m$, $\bar{r} = \mathbf{1}_\ell \otimes r$, and $\mathbf{A}_m \coloneqq I_\ell \otimes A_m$, $\mathbf{B}_m \coloneqq I_\ell \otimes B_m$ are Kronecker-structured matrices compatible with \eqref{E4}. The agent interactions governed by $\mathcal{G}_\ell$ and $\mathcal{G}_m$ are embedded in the Laplacian-like matrix $\mathbb{L}_\ell$ and the leader weight matrix $\mathbb{A}_m$.
To ensure tracking of the leader, the distributed control law for each agent is defined as:
\begin{subequations}
\begin{align}
    \bar{u}(t - \tau_u) &= \Theta^\top(t)\bar{\eta}(t) \label{E6a} \\
    &= \begin{bmatrix}
        \left[\theta_x^{1\top}(t),\ \theta_\zeta^{1\top}(t),\ \theta_r^1(t)\right] & \cdots & 0 \\
        \vdots & \ddots & \vdots \\
        0 & \cdots & \left[\theta_x^{\ell\top}(t),\ \theta_\zeta^{\ell\top}(t),\ \theta_r^\ell(t)\right]
    \end{bmatrix}
    \begin{bmatrix}
        \left[x_1^\top(t),\ x_1^\top(t-\tau_x),\ r(t - \tau_u)\right]^\top \\
        \vdots \\
        \left[x_\ell^\top(t),\ x_\ell^\top(t - \tau_x),\ r(t - \tau_u)\right]^\top
    \end{bmatrix} \label{E6b} \\
    &= \Phi_x^\top(t) \bar{x}(t) + \Phi_\zeta^\top(t) \bar{x}(t - \tau_x) + \Phi_r(t) \bar{r}(t - \tau_u) \label{E6c}
\end{align}
\end{subequations}
where the parameter matrices are structured as:
\[
\Phi_x = \diag\{\theta_x^1, \dots, \theta_x^\ell\}, \quad 
\Phi_\zeta = \diag\{\theta_\zeta^1, \dots, \theta_\zeta^\ell\}, \quad 
\Phi_r = \diag\{\theta_r^1, \dots, \theta_r^\ell\}.
\]

Substituting \eqref{E6c} into the system dynamics \eqref{E4}, and assuming ideal parameter matching such that $\bar{x} \equiv \bar{x}_m$, the following algebraic matching conditions must hold:
\begin{align}
    \mathbf{A} + \mathbf{B} \Phi_x^{\ast\top} = \mathbf{A}_m, \quad
    \mathbf{A}^\zeta + \mathbf{B} \Phi_\zeta^{\ast\top} = 0, \quad
    \mathbf{B} \Phi_r^\ast = \mathbf{B}_m \label{E7}
\end{align}
where $(\Phi_x^\ast, \Phi_\zeta^\ast, \Phi_r^\ast)$ denote the ideal control parameters. Under these conditions, the network synchronization objective is satisfied, i.e.,
\[
\lim_{t \to \infty} \bar{e}(t) = (\mathbb{L}_\ell \otimes I_n)\bar{x}(t) - (\mathbb{A}_m \otimes I_n)\bar{x}_m(t) \to 0.
\]

\begin{assumption}[Dynamics]\label{assu1}
    The dynamics of \eqref{E4} is unknown and the delays $(\tau_x,\tau_u)$ are known satisfying $\tau_x \leq \tau_u$. Also, we assume the signs of $\theta_r^\ast$ in $\Theta_i,\forall i=1\dots,\ell$ are known. Moreover, the external disturbances are assumed to be bounded and sufficiently less persistent than the reference signal $\bar{r}(t)$ and control input $\bar{u}(t)$. Finally, there exist matching control gains $(\Phi_x^\ast, \Phi_\zeta^\ast, \Phi_r^\ast)$ for any given communication network defined by $(\mathbb{L}_\ell, \mathbb{A}_m)$, satisfying the connectivity assumptions in Remarks~\ref{rem1} and \ref{rem2}.
\end{assumption}

\section{Adaptation Algorithms and Stability}\label{S3}
We consider the closed form of the tracking error $\bar{e} \coloneqq (\mathbb{L}_\ell\otimes I_n)\Bar{x} - (\mathbb{A}_m\otimes I_n)\Bar{x}_m$ to derive the adaptive control laws and analyze stability. Using the system dynamics in \eqref{E4}, \eqref{E5}, and the matching conditions in \eqref{E7}, the derivative of the error becomes:
\begin{subequations}
\begin{align}
    \dot{\bar{e}}(t) &= (\mathbb{L}_\ell\otimes I_n)\dot{\Bar{x}} - (\mathbb{A}_m\otimes I_n)\dot{\Bar{x}}_m \nonumber\\
    &= (\mathbb{L}_\ell\otimes I_n)\left[\mathbf{A}\Bar{x}(t) + \mathbf{A}^\zeta\Bar{x}(t-\tau_x) + \mathbf{B}\Bar{u}(t-\tau_u)\right] - (\mathbb{A}_m\otimes I_n)\left[\mathbf{A}_m\Bar{x}_m(t) + \mathbf{B}_m\Bar{r}(t-\tau_u)\right] \label{E8a}\\
    &= \left[(\mathbb{L}_\ell\otimes I_n)\mathbf{A}_m - (\mathbb{L}_\ell\otimes I_n)\mathbf{B}\Phi_x^{\ast\top}(t)\right]\Bar{x}(t) -(\mathbb{L}_\ell\otimes I_n)\mathbf{B}\Phi_\zeta^{\ast\top}(t)\Bar{x}(t-\tau_x) \nonumber \\
    &\quad + (\mathbb{L}_\ell\otimes I_n)\mathbf{B}\Bar{u}(t-\tau_u) - (\mathbb{L}_\ell\otimes I_n)\mathbf{A}_m\Bar{x}(t) + \mathbf{A}_m\Bar{e}(t) - (\mathbb{L}_\ell\otimes I_n)\mathbf{B}\Phi_r^\ast(t)\Bar{r}(t-\tau_u) \label{E8b}\\
    &= \mathbf{A}_m\Bar{e}(t) - (\mathbb{L}_\ell\otimes I_n)\mathbf{B}\left(\left[\Phi_x^{\ast\top}(t)\Bar{x}(t) + \Phi_\zeta^{\ast\top}(t)\Bar{x}(t-\tau_x) + \Phi_r^\ast(t)\Bar{r}(t-\tau_u)\right] + \Bar{u}(t-\tau_u)\right). 
    \label{E8c}
\end{align}
\end{subequations}
Introducing the parameter estimation error $\tilde{\Theta}(t)$ such that $\Theta(t) = \Theta^\ast + \tilde{\Theta}(t)$ and using the control law from \eqref{E6c}, we rewrite \eqref{E8c} as:
\begin{align}\begin{aligned}
    \dot{\bar{e}}(t) &= \mathbf{A}_m\Bar{e}(t) - (\mathbb{L}_\ell\otimes I_n)\mathbf{B}\Theta^{\ast\top}(t)\Bar{\eta}(t)  + (\mathbb{L}_\ell\otimes I_n)\mathbf{B}\Bar{u}(t-\tau_u) \\
    &= \mathbf{A}_m\Bar{e}(t) + (\mathbb{L}_\ell\otimes I_n)\mathbf{B}\left[\Tilde{\Theta}^{\top}(t) - \Theta^{\top}(t)\right]\Bar{\eta}(t)  + (\mathbb{L}_\ell\otimes I_n)\mathbf{B}\Bar{u}(t-\tau_u).
    \end{aligned}\label{E9}
\end{align}
However, this input-delay $\Bar{u}(t-\tau_u)$ should be adjusted for the sake of eliminating the required prediction to the control signal of $\Bar{u}(t+\tau_u|t)$. To tackle this issue, we replace it with a prediction based on the leader model $m$, defined as the time-varying function of $\Bar{\eta}_m(t)$, which is more easily to handle, leading to:
\begin{align}
    \Bar{u}(t) &= \Theta^\top(t)\Bar{\eta}_m(t+\tau_u|t). \label{E10}
\end{align}
Evaluating the delayed control input $\Bar{u}(t-\tau_u)$, it becomes the following:
\begin{align}
    \begin{aligned}
    \Bar{u}(t-\tau_u) &= \Theta^\top(t-\tau_u)\Bar{\eta}_m(t)\\
    &= \Phi_x^\top(t-\tau_u)\Bar{x}_m(t) + \Phi_\zeta^\top(t-\tau_u)\Bar{x}_m(t-\tau_x) + \Phi_r(t-\tau_u)\Bar{r}(t-\tau_u).
    \end{aligned} \label{E11}
\end{align}
Therefore, the error equation in \eqref{E9} is expressed as follows, 
\begin{align}\begin{aligned}
    \dot{\bar{e}}(t) &= \mathbf{A}_m\Bar{e}(t) + (\mathbb{L}_\ell\otimes I_n)\mathbf{B}\left[\Tilde{\Theta}^{\top}(t) - \Theta^{\top}(t)\right]\Bar{\eta}(t)  + (\mathbb{L}_\ell\otimes I_n)\mathbf{B}\Theta^{\top}(t-\tau_u)\Bar{\eta}_m(t) \\
    &= \mathbf{A}_m\Bar{e}(t) + (\mathbb{L}_\ell\otimes I_n)\mathbf{B}\left[\Tilde{\Theta}^{\top}(t)\Bar{\eta}(t) - \phi(t)\right]
    \end{aligned}\label{E12}
\end{align}
where the delay-term $\phi(t)$ is the inputs difference (\textit{mismatch between current and delayed inputs}) from two time-varying functions of $\Bar{\eta}(t)$ and $\Bar{\eta}_m(t)$ as follows,
\begin{align}
    \phi(t)\in\mathbb{R}^\ell = \Theta^{\top}(t)\Bar{\eta}(t) - \Theta^{\top}(t-\tau_u)\Bar{\eta}_m(t). \label{E13}
\end{align}
We now introduce an auxiliary gain matrix $\Phi_\phi = \diag\{\theta_\phi^1, \dots, \theta_\phi^\ell\}$ such that $\mathbf{B} = \mathbf{B}_m \Phi_\phi^\ast$. This allows us to reshape the error dynamics in \eqref{E12} as:
\begin{align}\begin{aligned}
    \dot{\bar{e}}(t) 
    &= \mathbf{A}_m\Bar{e}(t) + (\mathbb{L}_\ell\otimes I_n)\mathbf{B}\Tilde{\Theta}^{\top}(t)\Bar{\eta}(t) - (\mathbb{L}_\ell\otimes I_n)\mathbf{B}_m\Phi_\phi^\ast(t)\phi(t) \\
    &= \mathbf{A}_m\Bar{e}(t) + (\mathbb{L}_\ell\otimes I_n)\mathbf{B}\Tilde{\Theta}^{\top}(t)\Bar{\eta}(t) + (\mathbb{L}_\ell\otimes I_n)\mathbf{B}_m\Tilde{\Phi}_\phi(t)\phi(t) - (\mathbb{L}_\ell\otimes I_n)\mathbf{B}_m\Phi_\phi(t)\phi(t) 
    \end{aligned}
    \label{E14}
\end{align}
in which this $\Phi_\phi$ should be adjusted over time where $\Bar{u}_a = \Phi_\phi\phi$. To cancel the term involving $\Phi_\phi$, we define an auxiliary dynamic model as:
\begin{align}
    \dot{\Bar{x}}_a = \mathbf{A}_m\Bar{x}_a(t) + (\mathbb{L}_\ell\otimes I_n) \mathbf{B}_m\Phi_\phi(t)\phi(t). \label{E15}
\end{align}
\begin{remark}[Connectivity and Control Structure]
    If there exists any $i$ for which $\lambda_i^m = 0 < \vartheta$, then by Remark~\ref{rem2}, it follows that $\lambda_i^\ell = 0$, and thus $u_{a_i} = 0$, which implies $\theta_\phi^{\ast i} = 0$. This violates the requirement $B_i = B_m \theta_\phi^{\ast i}$ and invalidates the equivalence of $\mathbf{B}$ and $\mathbf{B}_m \Phi_\phi^\ast$. Therefore, the existence of a directed path from the leader to all agents is necessary, in line with the threshold assumption of $\vartheta > 0$ in Remark~\ref{rem1}.
\end{remark}
We define the augmented tracking error as:
\begin{align}
    \bar{e}_a(t) = \bar{e}(t) + \bar{x}_a(t) \label{E16}
\end{align}
and by combining \eqref{E14} and \eqref{E15}, the augmented error dynamics\footnote{The standard error dynamics in distributed adaptive control of networked system} becomes:
\begin{align}
    \dot{\Bar{e}}_a(t) = \mathbf{A}_m\Bar{e}_a(t) + (\mathbb{L}_\ell\otimes I_n)\mathbf{B}\Tilde{\Theta}^{\top}(t)\Bar{\eta}(t) + (\mathbb{L}_\ell\otimes I_n)\mathbf{B}_m\Tilde{\Phi}_\phi(t)\phi(t)
    \label{E17}
\end{align}
leading to natural form of adjustable parameterization. To ensure convergence $\lim_{t\to\infty}\Bar{e}_a = 0$ and $\lim_{t\to\infty}\Bar{x}_a = 0$, the adaptive parameter update laws of $\dot{\Theta}$ and $\dot{\Phi}_\phi$ are selected as:
\begin{align}
    \dot{\Tilde{\Theta}}^\top = \dot{\Theta}^\top &= -\sign(\Phi_r^\ast)\Gamma_\theta\mathbf{B}_m^\top(\mathbb{L}_\ell\otimes I_n)^\top P\Bar{e}_a(t)\Bar{\eta}^\top(t) \label{E18}\\
    \dot{\Tilde{\Phi}}_\phi = \dot{\Phi}_\phi &= -\Gamma_\phi\mathbf{B}_m^\top(\mathbb{L}_\ell\otimes I_n)^\top P\Bar{e}_a(t)\phi(t) \label{E19}
\end{align}
where $\Gamma_\theta, \Gamma_\phi \in \mathbb{R}^{\ell \times \ell}$ are symmetric positive-definite learning rate matrices, and $P = P^\top \succ 0 \in \mathbb{R}^{\bar{n} \times \bar{n}}$ is the solution to the Lyapunov equation:
\[
\mathbf{A}_m^\top P + P \mathbf{A}_m = - (Q + Q_a)
\]
with $Q = Q^\top \succ 0$ and $Q_a = Q_a^\top \succ 0$ being designer-defined positive definite matrices. To analyze the stability of the networked system under \eqref{E17}, we propose the following Lyapunov function candidate $V(\Bar{e}_a,\Tilde{\Theta},\Tilde{\Phi}_\phi)$:
\begin{subequations}
\begin{align}
    V(\Bar{e}_a,\Tilde{\Theta},\Tilde{\Phi}_\phi) &= \Bar{e}_a^\top(t)P\Bar{e}_a(t) + \int_{-\tau_x}^0\Bar{e}_a^\top(\sigma)Q_a\Bar{e}_a(\sigma)~d\sigma \label{E20a}\\
    &\quad + \tr\left[\Tilde{\Theta}(t)\Gamma_\theta^{-1}|\Phi_r^{\ast-1}|\Tilde{\Theta}^\top(t)\right] + \int_{-\tau_1}^0\tr\left[\Tilde{\Theta}(\sigma)\Gamma_\theta^{-1}|\Phi_r^{\ast-1}|\Tilde{\Theta}^\top(\sigma)\right]d\sigma \label{E20b}\\
    &\quad+ \tr\left[\Tilde{\Phi}_\phi(t)\Gamma_\phi^{-1}\Tilde{\Phi}_\phi^\top(t)\right] + \int_{-\tau_2}^0\tr\left[\Tilde{\Phi}_\phi(\sigma)\Gamma_\phi^{-1}\Tilde{\Phi}_\phi^\top(\sigma)\right]d\sigma. \label{E20c}
\end{align}
\end{subequations}
Taking the derivative of $V(\Bar{e}_a,\Tilde{\Theta},\Tilde{\Phi}_\phi)$ along the trajectories of \eqref{E17}, and using the property $\tr[ab^\top] = b^\top a$, we obtain:
\begin{subequations}
\begin{align}
    \dot{V} &= \Bar{e}_a^\top(t)\left[\mathbf{A}_m^\top P + P\mathbf{A}_m\right]\Bar{e}_a(t) + 2\Bar{e}_a^\top(t)P(\mathbb{L}_\ell\otimes I_n)\mathbf{B}\Tilde{\Theta}^{\top}(t)\Bar{\eta}(t) \nonumber\\
    &\quad+ 2\Bar{e}_a^\top(t)P(\mathbb{L}_\ell\otimes I_n)\mathbf{B}_m\Tilde{\Phi}_\phi(t)\phi(t) - \Bar{e}_a^\top(t-\tau_x)Q_a\Bar{e}_a(t-\tau_x) \nonumber\\
    &\quad+ 2\tr\left[\dot{\Tilde{\Theta}}(t)\Gamma_\theta^{-1}|\Phi_r^{\ast-1}|\Tilde{\Theta}^\top(t)\right] - \tr\left[\left(\Tilde{\Theta}(t) + \Tilde{\Theta}(t-\tau_1)\right)\Gamma_\theta^{-1}|\Phi_r^{\ast-1}|\left(\Tilde{\Theta}(t) + \Tilde{\Theta}(t-\tau_1)\right)^\top\right] \nonumber\\
    &\quad + 2\tr\left[\dot{\Tilde{\Phi}}_\phi(t)\Gamma_\phi^{-1}\Tilde{\Phi}_\phi^\top(t)\right] - \tr\left[\left(\Tilde{\Phi}_\phi(t) + \Tilde{\Phi}_\phi(t-\tau_2)\right)\Gamma_\phi^{-1}\left(\Tilde{\Phi}_\phi(t) + \Tilde{\Phi}_\phi(t-\tau_2)\right)^\top\right] \nonumber\\
    &= -\Bar{e}_a^\top(t)Q\Bar{e}_a(t)  - \tr\left[\left(\Tilde{\Phi}_\phi(t) + \Tilde{\Phi}_\phi(t-\tau_2)\right)\Gamma_\phi^{-1}\left(\Tilde{\Phi}_\phi(t) + \Tilde{\Phi}_\phi(t-\tau_2)\right)^\top\right] \nonumber\\
    &\quad - \tr\left[\left(\Tilde{\Theta}(t) + \Tilde{\Theta}(t-\tau_1)\right)\Gamma_\theta^{-1}|\Phi_r^{\ast-1}|\left(\Tilde{\Theta}(t) + \Tilde{\Theta}(t-\tau_1)\right)^\top\right] - \Bar{e}_a^\top(t-\tau_x)Q_a\Bar{e}_a(t-\tau_x)\nonumber\\
    &\quad+ 2\Bar{e}_a^\top(t)P(\mathbb{L}_\ell\otimes I_n)\mathbf{B}\Tilde{\Theta}^{\top}(t)\Bar{\eta}(t) - 2\tr\left[\Bar{\eta}(t)\Bar{e}_a^\top(t)P(\mathbb{L}_\ell\otimes I_n)\mathbf{B}_m\Gamma_\theta\Gamma_\theta^{-1}|\Phi_r^{\ast-1}|\Tilde{\Theta}^{\top}(t)\right] \label{E21a}\\
    &\quad+ 2\Bar{e}_a^\top(t)P(\mathbb{L}_\ell\otimes I_n)\mathbf{B}_m\Tilde{\Phi}_\phi(t)\phi(t) - 2\tr\left[\phi(t)\Bar{e}_a^\top(t)P(\mathbb{L}_\ell\otimes I_n)\mathbf{B}_m\Gamma_\phi\Gamma_\phi^{-1}\Tilde{\Phi}_\phi(t)\right] \label{E21b}\\
    &< 0. \label{E21c}
\end{align}
\end{subequations}
for appropriate selections of $\tau_1\leq\tau_x$ and $\tau_2\leq\tau_u$ and the designed $\dot{\Tilde{\Theta}}$ and $\dot{\Tilde{\Phi}}_\phi$ as \eqref{E18} and \eqref{E19} in turn. To end, we deliver the sufficient condition for stability in the following remark and the theorem covering the proposed idea. 
\begin{remark}[Stability Condition]
The non-integral part of the Lyapunov function in \eqref{E20a}-\eqref{E20c} is sufficient to guarantee stability and tracking. Define:
\begin{align}
    V_d(t) &= \bar{e}_a^\top(t) P \bar{e}_a(t)
    + \tr\left[ \tilde{\Theta}(t) \Gamma_\theta^{-1} |\Phi_r^{\ast-1}| \tilde{\Theta}^\top(t) \right]
    + \tr\left[ \tilde{\Phi}_\phi(t) \Gamma_\phi^{-1} \tilde{\Phi}_\phi^\top(t) \right], \label{E22}
\end{align}
then its derivative satisfies $\dot{V}_d = -\bar{e}_a^\top(t) Q \bar{e}_a(t) < 0$ due to the exact cancellation in \eqref{E21a}--\eqref{E21b}.
\end{remark}
\begin{theorem}[Stability and Tracking Guarantee]\label{L1}
    Consider the networked delayed system \eqref{E4} and the leader system \eqref{E5}, with the Laplacian-like matrix $(\mathbb{L}_\ell \otimes I_n)$ and leader weight matrix $(\mathbb{A}_m \otimes I_n)$ satisfying Remarks~\ref{rem1} and \ref{rem2}. Suppose the pair $(\mathbf{A}, \mathbf{B})$ is stabilizable and Assumption~\ref{assu1} holds. Let the control input be defined as
    \[
    \bar{u}(t) = \Theta^\top(t) \bar{\eta}_m(t + \tau_u | t)
    \]
    where $\eta_{m_i}:\mathbb{R}^+\to\mathbb{R}^{q}$, $\bar{\eta}_m = [\eta_{m_1}^\top, \dots, \eta_{m_\ell}^\top]^\top$ are known time-varying signals and $\Theta = \diag\{\Theta^1, \dots, \Theta^\ell\}$ with $\Theta^i = [\theta_x^{i\top}, \theta_\zeta^{i\top}, \theta_r^i]^\top$ be the adaptive term such that there exists the augmented model $\Bar{x}_a(t)$ in \eqref{E15} and the augmented error $\Bar{e}_a(t)$ in \eqref{E17}. Then, the adaptive update laws
    \begin{align}
    \begin{aligned}
        \dot{\tilde{\Theta}}^\top = \dot{\Theta}^\top &= -\sign(\Phi_r^\ast) \Gamma_\theta \mathbf{B}_m^\top (\mathbb{L}_\ell \otimes I_n)^\top P \bar{e}_a(t) \bar{\eta}^\top(t) \\
        \dot{\tilde{\Phi}}_\phi = \dot{\Phi}_\phi &= -\Gamma_\phi \mathbf{B}_m^\top (\mathbb{L}_\ell \otimes I_n)^\top P \bar{e}_a(t) \phi(t)
    \end{aligned}
    \label{E23}
    \end{align}
    ensure that the overall system is stable and that the tracking error satisfies $\lim_{t \to \infty} \bar{e}_a(t) = 0$. Hence, the equilibrium $(\bar{x}, \Theta)$ is uniformly stable.
\end{theorem}
\begin{remark}\label{rem5}
    Given \eqref{E4}, \eqref{E10}, \eqref{E20a}, and \eqref{E21c}, it follows from LaSalle’s Invariance Principle that the solution of \eqref{E4} asymptotically approaches the largest invariant set within $\mathbb{S}=\{\Bar{x}\in\mathbb{R}^{\bar{n}},:\dot{V}(\Bar{e}_a,\Tilde{\Theta},\Tilde{\Phi}_\phi)=0\}$.
\end{remark}

\paragraph{Summary.} Figure~\ref{F1c} illustrates the proposed framework. The control objective is to ensure that the agents governed by \eqref{E4} under network topology $\mathbb{L}_\ell$ track the leader defined in \eqref{E5} under influence of $\mathbb{A}_m$. To overcome state and input delays, the strategy employs predictor-based feedback derived from the leader and an auxiliary system, with adaptation laws provided in \eqref{E23}. Provided that the conditions in Remarks~\ref{rem1}--\ref{rem2} and Assumption~\ref{assu1} hold, Theorem~\ref{L1} guarantees convergence and stability.

\section{Numerical Results and Findings}\label{S4}
In this section, we present numerical simulations to validate the proposed distributed adaptive control framework for a network of uncertain agents under communication-induced delays. The simulations consider a network of $\ell = 4$ second-order agents, each subject to known state and input delays of $\tau_x = 3\,\text{s}$ and $\tau_u = 5\,\text{s}$, respectively. The agents' dynamics are defined with uncertain system matrices, while the leader model, serving as a reference, follows known stable dynamics, defined as follows:
\begin{gather*}
    \dot{\Bar{x}}_m(t) =  \left(I_4\otimes\begin{bmatrix}
        0 & 1 \\ -2 & -3
    \end{bmatrix}\right)\Bar{x}_m(t) + \left(I_4\otimes\begin{bmatrix}
        0 \\ -2
    \end{bmatrix}\right)\Bar{r}(t-\tau_u) \\
    \dot{\Bar{x}}(t) =  \diag\left\{\begin{bmatrix}
        0 & 1 \\ -2-i & -1-i
    \end{bmatrix}\right\}\Bar{x}(t) + \diag\left\{\begin{bmatrix}
        0 & 0 \\ \dfrac{2+i}{10} & \dfrac{2+i}{20}
    \end{bmatrix}\right\}\Bar{x}(t-\tau_x) + \diag\left\{\begin{bmatrix}
        0 \\ 2+i
    \end{bmatrix}\right\}\Bar{u}(t-\tau_u) 
\end{gather*}
for all $i=1,\dots,4$. Two different communication topologies are considered, corresponding to the networks shown in Figure~\ref{F1a} and Figure~\ref{F1b}. The associated Laplacian-like matrices $\mathbb{L}_\ell$ and leader weights $\mathbb{A}_m$ are selected to satisfy the connectivity condition $[(\mathbb{L}_\ell - \mathbb{A}_m)\otimes I_n]\mathbf{1}_{\bar{n}} = 0$:
\begin{align*}\begin{array}{cc|cc}
    \mathbb{L}_\ell = \begin{bmatrix}
        1 & 0 & 0 & 0 \\
        0 & 1 & 0 & 0 \\
        0 & 0 & 1 & 0 \\
        0 & 0 & 0 & 1
    \end{bmatrix}\otimes I_n, & \mathbb{A}_m = I_4\otimes I_n & \mathbb{L}_\ell = \begin{bmatrix}
        1 & -\gamma & 0 & -\gamma \\
        -\gamma & 1 & -\gamma & 0 \\
        0 & -\gamma & 1 & -\gamma \\
        -\gamma & 0 & -\gamma & 1
    \end{bmatrix}\otimes I_n, & \mathbb{A}_m = 0.4I_4\otimes I_n
    \end{array}
\end{align*}
for example 1 and 2 in turn where $\gamma=0.3$. Here, we design $\Gamma_\theta=\Gamma_\phi=I_4$, $Q+Q_a=I_4\otimes 0.1I_n$ while $P=I_4\otimes([p_1,p_2]/100)$ where $p_1 = [25~5]^\top, p_2 = [5~5]^\top$ satisfying $\mathbf{A}^\top_m P + P\mathbf{A}_m = -(Q+Q_a)$. Moreover, we design the initial conditions $\Theta(0)=-0.001\diag\{12.5,10,7.5,5\}\otimes\mathbf{1}_q$, $\Phi_\phi(0)=-0.1\diag\{4,3,2,1\}$.
\begin{figure}[h!]
    \centering
    \subfloat[\label{F2a}\small Tracking of states $x^i(t)$]{
        \includegraphics[width=0.3\textwidth]{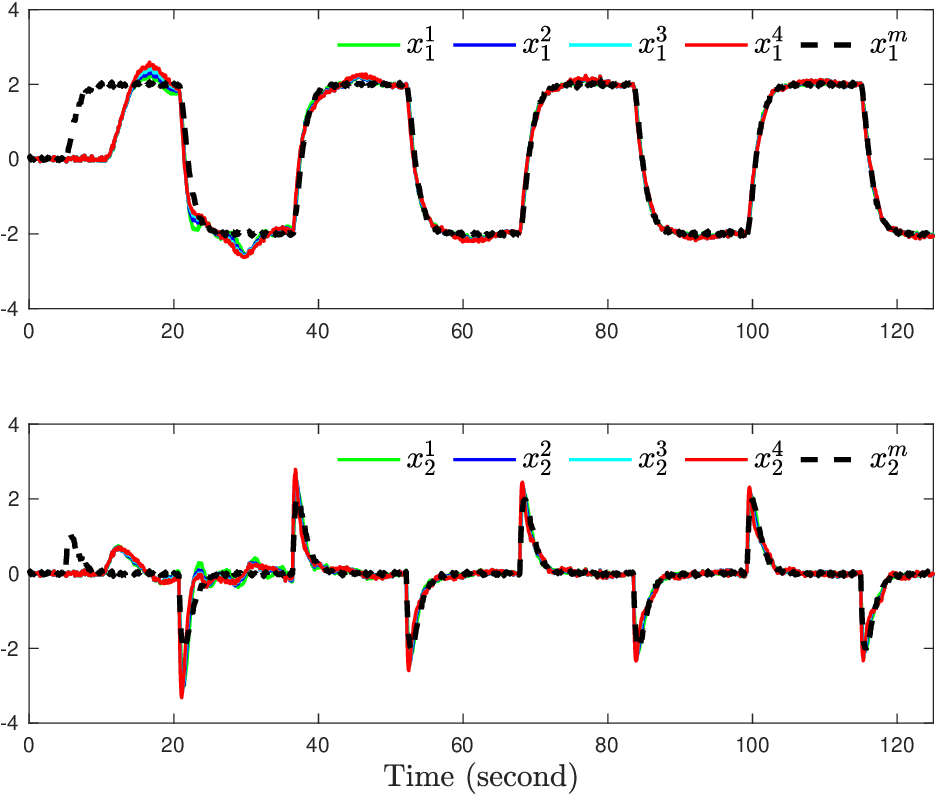}} \quad
    \subfloat[\label{F2b}\small $\Bar{r}(t),\Bar{u}(t),\Bar{u}_a(t)$]{
        \includegraphics[width=0.3\textwidth]{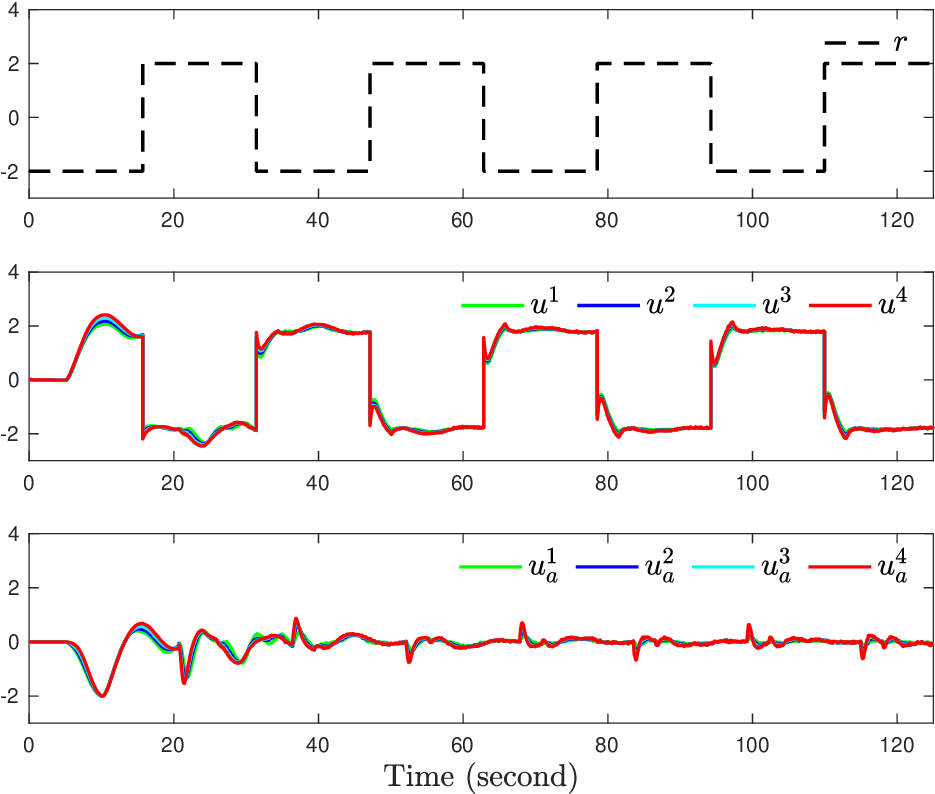}} \quad
    \subfloat[\label{F2c}\small Evolution of gains]{
        \includegraphics[width=0.3075\textwidth]{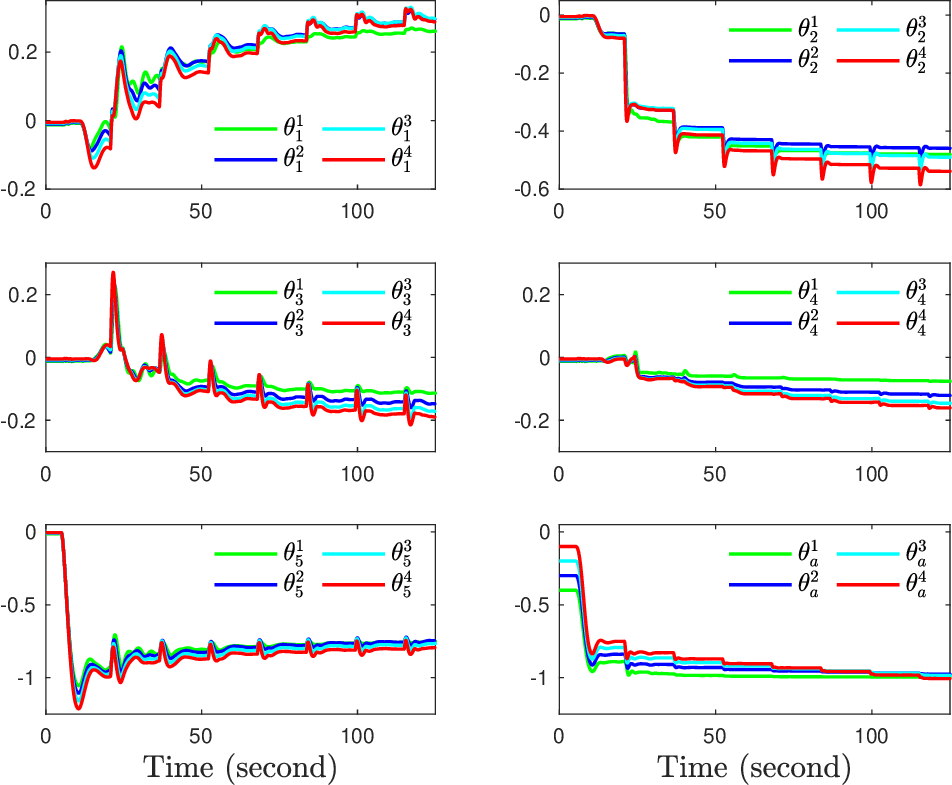}}\\ 
    \subfloat[\label{F2d}\small Tracking of states $x^i(t)$]{
        \includegraphics[width=0.3\textwidth]{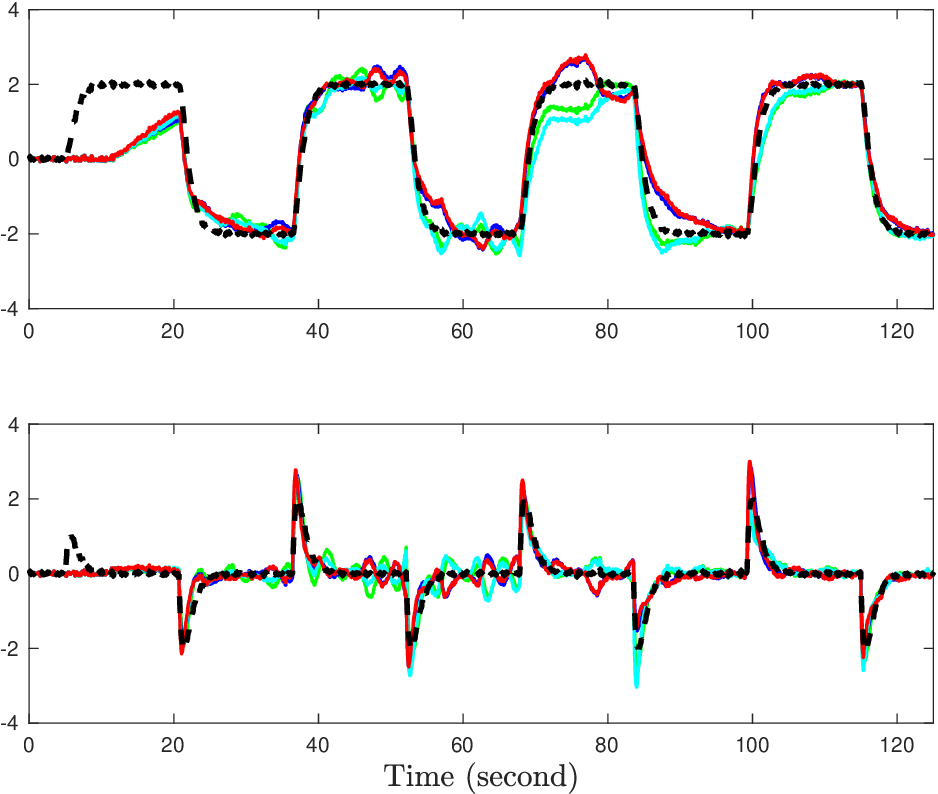}}\quad
    \subfloat[\label{F2e}\small $\Bar{r}(t),\Bar{u}(t),\Bar{u}_a(t)$]{
        \includegraphics[width=0.3\textwidth]{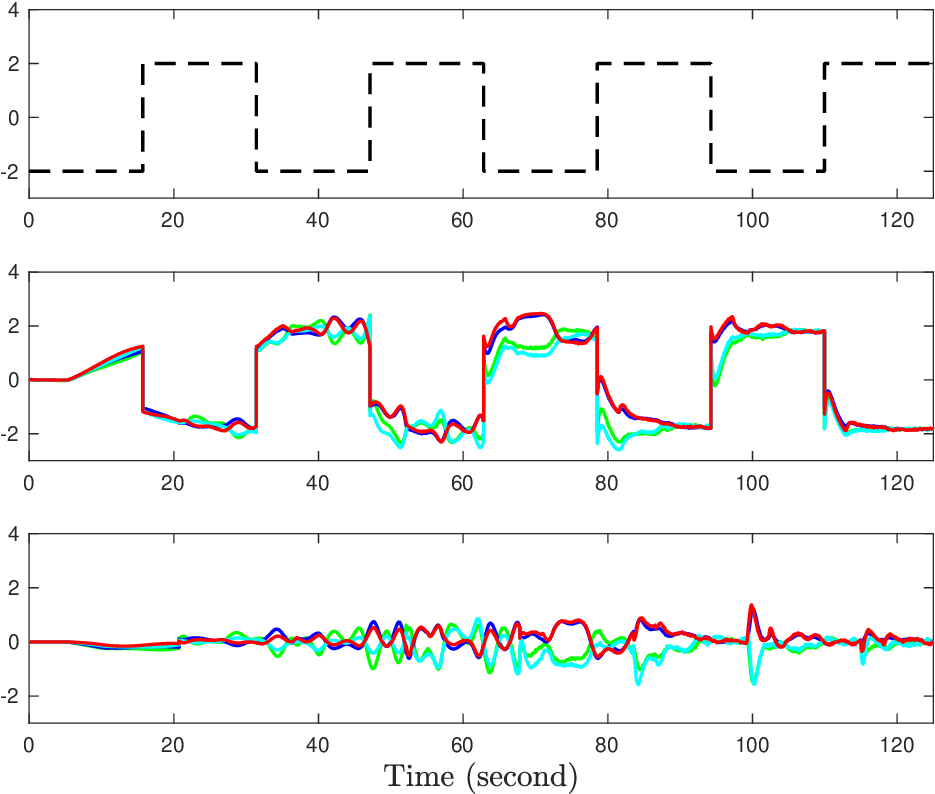}}\quad
    \subfloat[\label{F2f}\small Evolution of gains]{
        \includegraphics[width=0.3075\textwidth]{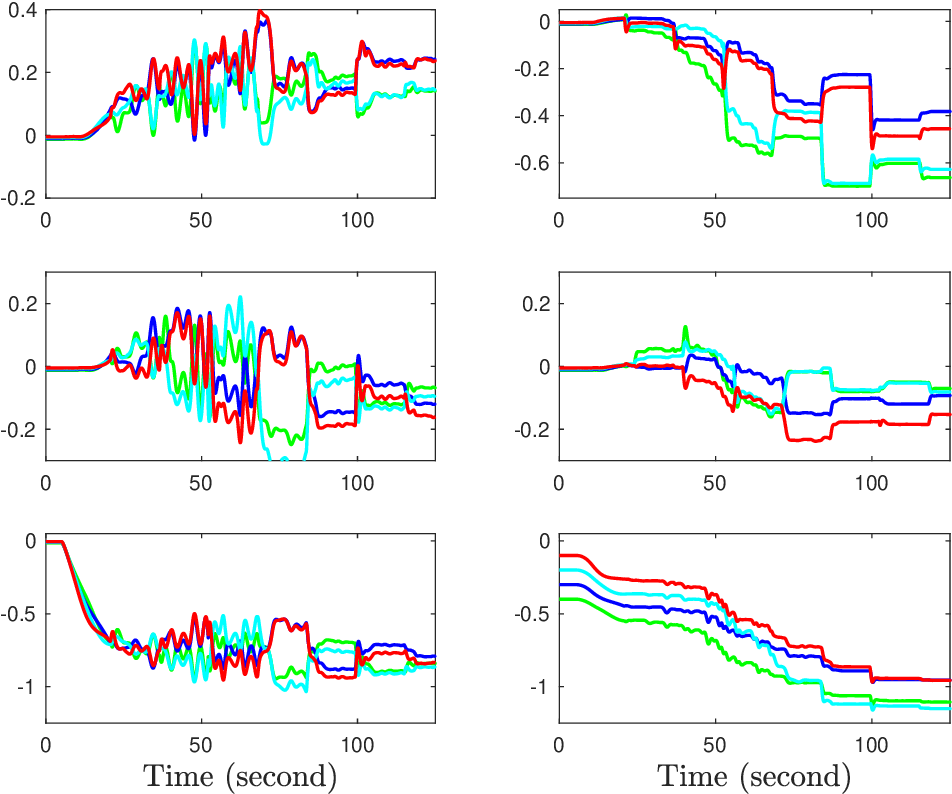}}
    \caption{(a), (b) and (c) are for example 1 in Figure~\ref{F1a} while (d), (e) and (f) are for example 2 in Figure~\ref{F1b}.}
    \label{F2}
\end{figure}

In the first scenario, illustrated in Figure~\ref{F1a}, each agent has direct access to the leader state with weight one, i.e., $w_{im}=1,\forall i$, and no communication among agents ($\mathbb{L}_\ell = I_4$). This setting ensures a high degree of leader influence and is expected to yield fast convergence. The results of this example are depicted in Figures~\ref{F2a}--\ref{F2c}. Figure~\ref{F2a} demonstrates excellent tracking performance, where each agent’s state quickly aligns with the leader’s trajectory. The tracking curves are nearly indistinguishable, indicating uniform and synchronized convergence. Figure~\ref{F2b} displays the evolution of the reference input $r(t)$, the distributed control input $\bar{u}(t)$, and the auxiliary control $\bar{u}_a(t)$. The auxiliary input effectively compensates for the delays and smooths the transient response, particularly around rapid changes in the reference. The evolution of the adaptive gains, shown in Figure~\ref{F2c}, exhibits monotonic and smooth convergence to steady-state values. This behavior is consistent with the theoretical predictions of the proposed Lyapunov-based adaptive law, confirming stable parameter adjustment.

In contrast, the second example considers a ring-type sparse network (Figure~\ref{F1b}) with limited direct leader influence, where $\mathbb{L}_\ell$ is structured with neighbor couplings weighted by $\gamma = 0.3$ and the leader weights are set to $\mathbb{A}_m = 0.4I_4\otimes I_n$. This configuration models a more distributed coordination structure with weaker global information, and thus more reliance on peer-to-peer interaction. The corresponding results are shown in Figures~\ref{F2d}--\ref{F2f}. In Figure~\ref{F2d}, the agents still manage to track the leader state, but the convergence is noticeably slower, and transient fluctuations are more pronounced. The increased sensitivity to initial conditions and delay effects is evident. Figure~\ref{F2e} shows greater oscillatory behavior in both the control input and auxiliary compensation, reflecting the greater complexity in coordinating among agents without full leader access. Nevertheless, despite the less favorable network conditions, the adaptive gains illustrated in Figure~\ref{F2f} still converge to bounded values over time. Although their trajectories are less smooth than in the fully connected case, they exhibit stability and eventual saturation, demonstrating the robustness of the proposed adaptation mechanism.

Overall, the simulations confirm the theoretical guarantees of the proposed control scheme. Example 1, with full leader connectivity, achieves rapid and stable convergence due to direct access to the reference dynamics. Example 2, while slower and more oscillatory, still achieves perfect tracking asymptotically, thanks to the adaptive nature of the distributed control law. The auxiliary input and gain adaptation laws effectively compensate for the input delays and inter-agent communication limitations. These results collectively highlight the flexibility and scalability of the proposed method in handling uncertain, delayed, and distributed multi-agent systems.

\section{Conclusion}\label{S5}
This paper presented a comprehensive adaptive control framework for heterogeneous multi-agent systems with unknown dynamics and communication-induced delays. A leader-follower structure was employed, where each agent seeks to track a reference trajectory generated by a known leader system. The theoretical development included a detailed formulation of the problem, the construction of distributed adaptive control laws, and a rigorous stability analysis using Lyapunov methods.

Key elements of the framework—such as the network connectivity threshold and the balanced communication topology—were highlighted through formal remarks, establishing the boundaries within which the proposed solution is guaranteed to operate. Importantly, the design effectively compensates for both state and input delays, including their lumped effects, enabling all agents to asymptotically synchronize with the leader despite uncertainty and delayed information flow.

Numerical simulations were conducted on two representative network configurations to demonstrate the efficacy of the proposed control strategy. In both examples, the agents successfully tracked the reference model, with variations in convergence behavior attributable to the network structure. The results confirmed that the method achieves perfect tracking under varying levels of connectivity and communication sparsity.

Future work will focus on extending this approach to address resilience in the presence of faults or attacks. This includes incorporating learning-based mechanisms for fault detection and reconfiguration in distributed adaptive control of networked systems, as motivated in recent studies~\cite{R44,R45}.

\section*{Acknowledgement}\label{}
This research is fully supported by Department of Engineering Physics, Institut Teknologi Sepuluh Nopember, under grant No.1868/PKS/ITS/2023

\bibliographystyle{IEEEtran}    
\bibliography{Referee}

\end{document}